\documentclass[doublecol]{epl2}
\usepackage{graphicx}
\usepackage{cite}
\usepackage{amsmath}
\usepackage{amsfonts}
\usepackage{amssymb}
\usepackage{color}
\usepackage{float}
\usepackage{multirow,multicol}
\usepackage{tabulary}
\usepackage[flushleft]{threeparttable}
\usepackage{xcolor}
\usepackage{ulem}
\usepackage{hyperref}
\usepackage{nameref}

\title{Electric quadrupole moment of a neutral non-relativistic particle in the presence of screw dislocation}
\shorttitle{Title} 

\author{H. Hassanabadi\inst{1,2} \and S. Zare\inst{1} \and J. K\v{r}\'i\v{z}\inst{2}  \and B.C. L\" utf\"uo\u{g}lu\inst{2,3}  }
\shortauthor{H. Hassanabadi \etal}

\institute{
  \inst{1} Faculty of Physics, Shahrood University of Technology, Shahrood, Iran.\\
  \inst{2} Department of Physics, University of Hradec Kr\'alov\'e,
				Rokitansk\'eho 62, 500 03 Hradec Kr\'alov\'e, Czechia. \\
  \inst{3} Department of Physics, Akdeniz University, Campus 07058 Antalya, Turkey.
}
\pacs{03.65.Ge}{Solutions of wave equations$:$ bound states}
\pacs{03.65.Ta}{Foundations of quantum mechanics$;$ measurement theory}
\pacs{03.65.−w}{Quantum mechanics}

\abstract{
In this contribution, we investigate the interaction between electric and magnetic fields with an electric quadrupole moment of a spinless particle moving in an elastic medium which has a topological defect (screw dislocation).  By considering this interaction, the Schr\"odinger equation is exactly solved by using analytical method. Thus, the eigenfunction and energy eigenvalues for two configurations is found. Meanwhile, by observing a shift in the angular momentum quantum number,  the energy eigenvalues and the wave function of the system are modified, due to the screw dislocation in the medium.}
\DeclareUnicodeCharacter{2212}{-}
\begin{document}

\maketitle
\section{Introduction}
The universe must have exceeded the critical temperature at one time such that at first, the symmetry was unbroken; then, it expands and cools. Thus, the universe could acquire the topological defects arising from the symmetry-breaking phase transitions  \cite{Kibble1976,Vilenkin1994,Vilenkin1985,Bunkov2012,Puntigam,Carvalho2011,WangEPL2019}, according to the {Big-Bang} model. Thus, the topological defects that have attracted attention in some fields of physics could be formed by phase transitions during the universe evolution. The topological defects appear as linear defects (dislocations and disclinations) \cite{Puntigam,Nabarro,ZareEPJP2020,Kleinert,Dzyaloshinskii,Bausch,Katanaev} in the condensed-matter physics, also they have investigated by cosmology, that is,  cosmic strings and monopoles in cosmology. It worth mentioning that the equivalence between the theory of defects in three-dimensional gravity and solids with torsion has been vouched by Katanaev and Volovich \cite{Katanaev}. According to this geometric theory of defects \cite{Katanaev2005,DantasPLA2015}, the topological defect causes elastic deformations in continuous media so that these elastic deformations are described by a line element (metric). In fact, the continuous elastic medium is characterized by a Riemann–Cartan manifold.  In this medium, torsion and curvature \cite{Bilby1955} are related to dislocations (screw dislocation and spiral dislocation) and disclinations, respectively. Over the past few decades, a considerable interest in screw dislocation was observed, so that the influences of screw dislocation on quantum systems have been studied \cite{deAMarques2001,FurtadoPLA2001,FurtadoJPA2000,FilgueirasPLA2015,BakkePB2018}.\\
The interaction between multipole moments and electric and magnetic fields in the context of the quantum dynamics can give rise to striking quantum effects, such as the appearance of quantum topological phase \cite{Aharonov1959,Wilkens1994,He1993,Peshkin,Anandan1989,AndradePRD2012,BakkeAP2014} and the Landau quantization \cite{LemosdeMelo2011}. Beside, the quantum dynamics of electric quadrupole moment
has been studies, For example, Chen \cite{ChenPRA1995}
showed that the Aharonov-Bohm effect is a generic feature of any multipole moment moving in an electromagnetic field, also as explicit examples, the electric and magnetic quadrupole moments are discussed. Lemos de Melo et al. \cite{LemosdeMelo2011} proposed a model for achieving the Landau quantization for an electric quadrupole moment. Bakke \cite{Bakke2012} showed that, by choosing a diagonal electric quadrupole tensor whose components are constants, the interaction between an electric quadrupole moment and electric fields gives rise to the scalar Aharonov-Bohm effect and to bound states analogous to the confinement of a quantum particle to a quantum dot. \\
In this study, we are investigate the interaction between electric and magnetic fields with {an} electric quadrupole moment of a spinless particle moving in an elastic medium which has a screw dislocation.  By considering this interaction, the Schrodinger equation is exactly solved by using analytical method. Thus, the eigenfunction and energy eigenvalues for two configurations is found. Beside, we can observe the shift in the angular momentum quantum number which affects the energy eigenvalues and the wave function of the system, due to the screw dislocation in medium. \\
This paper is organized as follows. { In the following section} a concise review of the classical dynamics for an electric quadrupole moment is presented, and a non-relativistic moving particle that has an electric quadrupole moment is considered in an elastic medium with a screw dislocation in the presence of an electric and magnetic magnetic fields. Accordingly, the second-order radial Schr\"odinger equation is obtained.
{ Then, in the next section,} the Schr\"odinger equation is exactly solved and acquired the eigenfunction and energy eigenvalues for two cases, the investigation of interaction in the absence of potential and in the presence of a static scalar potential by using the Nikiforov-Uvarov (NU) method. { In the final section, } conclusions are presented.

\section{Electric quadrupole moment of a moving particle in the presence of a screw dislocation\label{sec2}}

We begin our study of the influence of a topological defect in quantum dynamics for a spinless particle moving with an electric quadrupole moment interacting with electric and magnetic fields in an elastic medium in the presence of a screw dislocation. According to the geometric theory of topological defects,  the screw dislocation is considered a model of the linear topological defect in an elastic medium. Thus,  in cylindrical coordinates, the screw dislocation along the $z$-axis is described by the following line element \cite{DantasPLA2015,FurtadoEPL1999,Netto}:
\begin{equation}\label{le1}
\mathrm{d}s^2=-\mathrm{d}t^2+\mathrm{d}\rho^2+\left(\mathrm{d}z+\beta\mathrm{d}\varphi\right)^2+\rho^2\mathrm{d}\varphi^2,
\end{equation}
where, the time coordinate is specified by $0<t<\infty$ and the spatial coordinates are determined by $\rho>0$ (that is the radial coordinate $\rho=\sqrt{x^2+y^2}$), $0<\varphi<2\pi$ (that is the azimuthal coordinate), and $-\infty<z<\infty$ (that is the vertical coordinate). The units is used as $\hbar=\mathrm{c}=1$. The parameter $\beta$ is a constant associated with the component of Burgers vector $\vec{b}=\left(0,0,b_{z}\right)$ so that it becomes $\beta=\frac{\mathrm{b}_{z}}{2\pi}$.
In fact, the constant parameter $\beta$ determines the topological defect (screw dislocation). With respect to Eq. \eqref{le1}, this kind of dislocation presents a torsion field that corresponds to a conical singularity at the origin (as mentioned in Ref. \cite{Netto}). The covariant metric tensor $g_{ij}$ and contravariant metric tensors $g^{ij}$ associated with the spatial part of the metric Eq. \eqref{le1} can be written as
\begin{equation}\label{mettens}
\mathrm{g}_{ij}=\begin{pmatrix}
1&0&0\\
0&\rho^2+\beta^2&\beta\\
0&\beta&1
\end{pmatrix}, \quad \mathrm{g}^{ij}=\begin{pmatrix}
1&0&0\\
0&\frac{1}{\rho^2}&-\frac{\beta}{\rho^2}\\
0&{-}\frac{\beta}{\rho^2}&1+\frac{\beta^2}{\rho^2}
\end{pmatrix}.
\end{equation}
Then, the
Laplace-Beltrami operator in the presence of a topological defect (which corresponds to the metric Eq. \eqref{le1}) can be written as \cite{deAMarques2001,FurtadoEPL1999}
\begin{equation}\label{Lap1}
\nabla^2=\frac{1}{\sqrt{g}}\partial_{i}\left(\sqrt{g}g^{ij}\partial_{j}\right),
\end{equation}
where $g=\mathrm{det}\left|g_{ij}\right|$, the Laplacian operator Eq. \eqref{Lap1} can be found as
\begin{equation}\label{Lap2}
\nabla^2=\frac{1}{\rho}\partial_{\rho}\left(\rho\partial_{\rho}\right)+\frac{1}{\rho^2}\left(\partial_{\varphi}-\beta\partial_{z}\right)^2+\partial_{z}^2,
\end{equation}
where
\begin{equation}\label{comgrad}
\nabla_{\varphi}=\frac{1}{\rho}\left(\partial_{\varphi}-\beta\partial_{z}\right),
\end{equation}
consequently, we can observe $\partial_{\varphi}\rightarrow\partial_{\varphi}-\beta\partial_{z}$ \cite{DantasPLA2015,Netto,SilvaEPJC2019}, according to the torsion (screw dislocation) in the elastic medium which is described by the line element Eq. \eqref{le1}.\\
Let us continue this study with a concise review of the classical dynamics for an electric quadrupole moment. Thus, in the rest frame of a particle, the potential energy of a multipole expansion \cite{BakkeAP2014,LemosdeMelo2011,ChenPRA1995,Bakke2012,BakkeIJMPA2014} can be denoted by
\begin{equation}\label{enerpot1}
U=q\Phi-\vec{d} {\cdot} \vec{\nabla}\Phi+\sum_{ij}Q_{ij}\partial_{i}\partial_{j}\Phi\dots,
\end{equation}
in which, the electric charge is denoted by $q$, the electric potential is specified by $\Phi$, such that the electric field $\vec{E}$  in the laboratory frame can be found by $\vec{E}=-\vec{\nabla}\Phi$, the electric dipole moment is determined by $\vec{d}$, the electric quadrupole moment tensor is denoted by $Q_{ij}$. In this contribution, we are interested in investigating the dynamics of an electric quadrupole moment; thereby we must suppose $q=0$ and $\vec{d}=0$ in Eq. \eqref{enerpot1}; accordingly, we can rewrite Eq. \eqref{enerpot1} as follows
\begin{equation}\label{enerpot2}
U=-\sum_{ij}Q_{ij}\partial_{i}E_{j}.
\end{equation}
For a moving particle which it possesses an electric quadrupole moment,  we have the fact that the mentioned particle interacts with a different electric field $\vec{E}'$ where $\vec{E}'=\vec{E}+\vec{v}\times\vec{B}$
(up to O($v^2$);
moreover, the velocity vector of the moving patricle is given by $\vec{v}$); meanwhile, the magnetic fields in the laboratory frame is denoted by $\vec{B}$. It should be noted that we must consider the Lorentz transformation of the electromagnetic field, such that we must substitute the electric field $\vec{E}$ in Eq. \eqref{enerpot2} with the electric field $\vec{E}'$ \cite{ChenPRA1995,LemosdeMelo2011,Bakke2012}. Thus, the
Lagrangian of this system can be written as
\begin{equation}\label{Lagrangian}
L=\frac12 m v^2+\vec{Q}.\vec{E}-\vec{v}.\left(\vec{Q}\times\vec{B}\right),
\end{equation}
in which, the $m$ is mass of moving particle, and the $\vec{Q}$ is a vector whose components are $\vec{Q}_{i}$ so that the components $\vec{Q}_{i}$ are defined as $\vec{Q}_{i}=\sum_{ij}Q_{ij}\partial_{j}$. Furthermore, the tensor  $Q_{ij}$ is a symmetric and traceless tensor (as mentioned in Refs. \cite{ChenPRA1995,LemosdeMelo2011,Bakke2012}). Then,  as regards to Eq. \eqref{Lagrangian}, let us write the canonical momentum as $\vec{p}=m \vec{v}-(\vec{Q}\times\vec{B})$. Thereby, the Hamiltonian of this system in the frame of a spinless particle moving with an electric quadrupole moment is given by
\begin{equation}\label{Hamilt}
H=\frac{1}{2m}\left[\vec{p}+\left(\vec{Q}\times\vec{B}\right)\right]^2-\vec{Q} \cdot \vec{E}+V\left(\rho\right).
\end{equation}
The common method to present the coupling between a charged particle and electromagnetic fields in the quantum mechanics is via the minimal coupling. According to this method, the differential operator $\vec{p}=-i\vec{\nabla}$ is remodelled by finding another contribution to this Hermitian operator, that is, adding the electromagnetic vector potential $\vec{A}$, such that the operator $\vec{p}=-i\vec{\nabla}$ can be modofied as  $\vec{p}\rightarrow-i\vec{\nabla}- q \vec{A}$ \cite{Medeirosa2010}. In this case, because our focus lies on the dynamics of an electric quadrupole moment, we set $q=0$ and $d=0$. In this way, we vanish the contribution associated with the electromagnetic vector potential $\vec{A}$ in the Hermitian operator. Accordingly, we should not consider $(-\vec{Q}\times\vec{B})$ to be similar to  $\vec{A}_{\mathrm{eff}}(\rho)$. On the other hand, the second term of the right-hand side
of Eq. \eqref{Hamilt}, that is, the term $(-\vec{Q} \cdot \vec{E})$ can be considered as a  scalar potential. Therefore, we suppose $V_{\mathrm{eff}}(\rho)=-\vec{Q} \cdot \vec{E}$. As regards to Eq. \eqref{Hamilt}, in non-relativistic systems, let us now present the Schr\"odinger equation for a moving particle with an electric quadrupole moment interacting with electric and magnetic fields in the presence of a screw dislocation in an elastic medium as
\begin{equation}\label{Schr1}
\left(\frac{\hat{\pi}^2}{2m}+V_{\mathrm{eff}}(\rho)+V(\rho)\right)\Psi\left(t,\vec{r}\right)=i\partial_{t}\Psi\left(t,\vec{r}\right),
\end{equation}
where $\hat{\pi}$ is generalized momentum operator given by
\begin{equation}
\hat{\pi}=-i\vec{\nabla}+(\vec{Q}\times\vec{B}).
\end{equation}
To find the second-order radial Schr\"odinger equation associated with Eq. \eqref{Schr1}, we can take into account a special case in which the non-null components of electric quadrupole moment tensor $Q_{ij}$ \cite{Bakke2012} are
\begin{equation}
Q_{\rho\rho}=Q_{\varphi\varphi}=-Q,\quad Q_{zz}=2Q,\qquad Q>0,
\end{equation}
in which $Q$ is a constant. Let us now proceed by considering a field configuration (as proposed in Ref. \cite{Bakke2012}) and in the elastic medium as follows
\begin{equation}\label{fields}
\vec{E}=\frac{\lambda}{\rho}\hat{\rho},\qquad \vec{B}=\frac{C_{m}}{2} \rho^2 \hat{z},
\end{equation}
in which $\lambda$ is considered as a linear electric charge density alonge the $z$-axis, and $C_{m}$ is considered as a constant. According to electric field Eq. \eqref{fields}, the electric potential can be acquired as $\Phi=-\lambda\,\mathrm{ln}\rho$.

With regard to Eq. \eqref{Schr1}, we now proceed our process by obtaining the operator $\hat{\pi}^2\Psi\left(t,\vec{r}\right)$ as follows
\small
\begin{equation}\label{Oppi1}
\hat{\pi}^2\Psi\left(t,\vec{r}\right)=\left[-\nabla^{2}-2i\left(\vec{Q}\times\vec{B}\right) \cdot \vec{\nabla}+\left(\vec{Q}\times\vec{B}\right)^2\right]\Psi\left(t,\vec{r}\right),\,\,\,
\end{equation}
\normalsize
thus, by substituting Eqs. \eqref{Lap2} and \eqref{comgrad}, and  also by using $\vec{Q}\times\vec{B}=Q C_{m}\rho \hat{\varphi}$, we can rewrite Eq. \eqref{Oppi1} as
\begin{eqnarray}\label{Oppi2}
\hat{\pi}^2\Psi\left(t,\vec{r}\right)&=&-\bigg[\frac{1}{\rho}\partial_{\rho}\left(\rho\partial_{\rho}\right)+\frac{1}{\rho^2}\left(\partial_{\varphi}-\beta\partial_{z}\right)^2+\partial_{z}^2 \nonumber\\
&+&2iQC_{m}\left(\partial_{\varphi}-\beta\partial_{z}\right)-Q^2C_{m}^2\rho^2
\bigg]\Psi\left(t,\vec{r}\right).\,\,\,\,\,\,\,\,\,\,\,
\end{eqnarray}
We now continue our analyze by choosing a special solution to Eq. \eqref{Schr1} as
\begin{equation}\label{WF1}
\Psi\left(t,\vec{r}\right)=e^{-i\mathcal{E}t+i\ell\varphi+ikz}\psi\left(\rho\right),
\end{equation}
this choice Eq. \eqref{WF1} leads to having a radial Schr\"odinger equation. It should be noted that this choice Eq. \eqref{WF1} is a common eigenfunction to operators Hamiltonian, $\hat{\mathrm{p}}_{z}$, and $\hat{\mathrm{L}}_{z}$ (the Hamiltonian commutes with the operators $\hat{\mathrm{p}}_{z}$  and $\hat{\mathrm{L}}_{z}$). As regards to Eq. \eqref{WF1}, the energy of this system is indicated by $\mathcal{E}$, the wave number (as a constant) along the $z$-axis is indicated by $k$  and the angular momentum quantum number in denoted by $\ell=0,\pm1,\pm2,\dots$. Thereby, by replacing eigenfunction Eq. \eqref{WF1} in Eq. \eqref{Schr1} and by considering $V_{\mathrm{eff}}=-\frac{Q\lambda}{2}$, we acquire the second-order radial Schr\"odinger equation as
\begin{equation}\label{Schr2}
\begin{split}
&\left[\frac{\mathrm{d}}{\mathrm{d}\rho^2}+\frac{1}{\rho}\frac{\mathrm{d}}{\mathrm{d}\rho}-\frac{1}{\rho^2}\left(\ell-\beta k\right)^2-k^2-2QC_{m}\left(\ell-\beta k\right)\right.\\
&\left.-Q^2C_{m}^2\rho^2+\frac{2mQ\lambda}{\rho^2}+2m\mathcal{E}-2m\mathrm{V}\left(\rho\right)\right]\psi\left(\rho\right)=0.
\end{split}
\end{equation}


\section{Exact solution of schr\"odinger equation\label{sec3}}
In this section, the radial Schr\"odinger equation which is acquired for a moving particle under a field configuration in an elastic medium with a torsion associated with a topological defect (screw dislocation) corresponds to a conical singularity at the origin is exactly solved and found the eigenfunction and energy eigenvalues for two cases, in the first case, we investigate the interaction in the absence of potential, and in the second case, we examine the interaction in the presence of a static scalar potential by using the Nikiforov-Uvarov (NU) method. Thus, let us start the first case as follows:

\subsection{\bf{The first case}}
In this case, we investigate our configuration analytically in the absence of potential, that is, $V(\rho)=0$. Thus, we can rewrite Eq. \eqref{Schr2} as
\small
\begin{eqnarray}\label{Schr3}
&&\frac{\mathrm{d}^2\psi_{n\ell}\left(\rho\right)}{\mathrm{d}\rho^2}+\frac{1}{\rho}\frac{\mathrm{d}\psi_{n\ell}\left(\rho\right)}{\mathrm{d}\rho}
 \nonumber \\
&+&\frac{1}{\rho^2}\Big[-Q^2C_{m}^2\rho^4+\left(2m\mathcal{E}-k^2-2QC_{m}\left(\ell-\beta k\right)\right)\rho^2\nonumber \\
&-&\left(\left(\ell-\beta k\right)^2-2mQ\lambda\right)\Big]\psi_{n\ell}\left(\rho\right)=0.
\end{eqnarray}
\normalsize
By choosing a changing variable as $s=\rho^2$, the Eq. \eqref{Schr3} can be rewritten as follows
\small
\begin{eqnarray}\label{Schr4}
&&\frac{\mathrm{d}^2\psi_{n\ell}\left(s\right)}{\mathrm{d}s^2}+\frac{1}{s}\frac{\mathrm{d}\psi_{n\ell}\left(s\right)}{\mathrm{d}s} \nonumber \\
&+&
\frac{1}{4s^2}\Big[-Q^2C_{m}^2s^2+\left(2m\mathcal{E}-k^2-2QC_{m}\left(\ell-\beta k\right)\right)s\nonumber \\ &-&\left(\left(\ell-\beta k\right)^2-2mQ\lambda\right)\Big] \psi_{n\ell}\left(s\right)=0.
\end{eqnarray}
\normalsize
According to the Appendix in Ref. \cite{deMontignyGRG2018}, we can observe that Eq. \eqref{Schr4}  is similar to the NU equation form \cite{ZhangPLA2010,SHDongPLA2005}. Thus, with regard to the eigenfunction form associated with NU equation, we can present the eigenfunction correspond to the generalized Laguerre polynomials $L^{\alpha}_{n}(x)$ as follows
\begin{equation}\label{WF2}
\psi_{n\ell}\left(\rho\right)=N\rho^{2\alpha_{12}}e^{\alpha_{13}\rho^2}L^{\alpha_{10}-1}_{n}\left(\alpha_{11}\rho^2\right),
\end{equation}
in which, the normalization constant is demonstrated by $N$, and also $\alpha_{10}$, $\alpha_{11}$, $\alpha_{12}$ and $\alpha_{13}$ is given by
\begin{eqnarray}\label{alphas1}
\alpha_{10}&=&1+\sqrt{\left(\ell-\beta k\right)^2-2mQ\lambda}, \nonumber \\ \alpha_{11}&=&\left|QC_{m}\right|, \nonumber \\
\alpha_{12}&=&\frac12\sqrt{\left(\ell-\beta k\right)^2-2mQ\lambda}, \nonumber \\ \alpha_{13}&=&-\frac12\left|QC_{m}\right|.
\end{eqnarray}
Then, the energy eigenvalues of this system can be written as
\begin{equation}\label{energyc1}
\mathcal{E}_{n\ell}=\frac{k^2}{2m}+\frac{QC_{m}}{m}\left(\ell-\beta k\right)+\frac{\delta}{m}\left|QC_{m}\right|.
\end{equation}
Where
\begin{equation}\label{delta1}
\delta=\left[1+2n+\sqrt{\left(\ell-\beta k\right)^2-2mQ\lambda}\right]
\end{equation}
Now, by focusing on the Eq. \eqref{energyc1}, we can observe that the energy eigenvalue subject to a given collection of parameters involving $n$ and $\ell$ (those are the quantum numbers), $m$ (that is mass of particle), $Q$ (that is a constant associated with the non-null components of tensor $Q_{ij}$), $\lambda$ (that is a linear electric charge density), $C_{m}$ (that is a constant related to the magnetic field), $k$ (that is the wave number) and $\beta$ (that is a constant associated with the screw dislocation). Considering $\left(\ell-\beta k\right)^2>2mQ\lambda$ leads to having a positive quantity in Eq. \eqref{delta1}.
Accordingly, the energy eigenvalues Eq. \eqref{energyc1} is clearly unbound state if $Q$ and $C_{m}$ have the same sign and also
$\ell>\beta k$.


\subsection{\bf{The second case}}
\begin{figure}[htb]
\onefigure[scale=0.55]{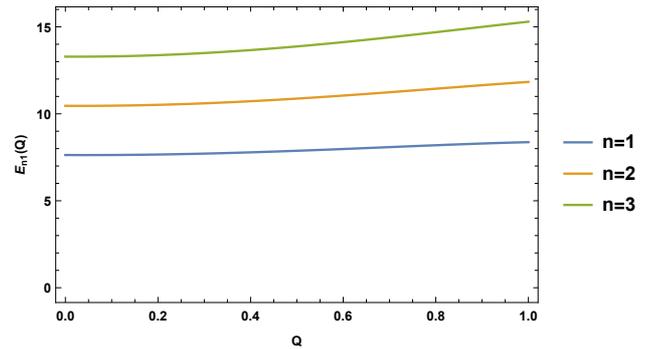}
\caption{The energy eigenvalue $\mathcal{E}_{n\ell}$ from Eq. \eqref{energyc2}, as a function of $Q$ for the values $\ell=\lambda=m=C_{m}=C_{1}=C_{2}=C_{3}=1$, with $k=0.5$ and $\beta=0.5$, according to three different values $\mathrm{n}=1,2,3$.}
\label{Fig1}
\end{figure}

In this case, we investigate our configuration analytically in the presence of a static scalar potential a follows
\begin{equation}\label{sspot}
\mathrm{V}\left(\rho\right)=C_1\rho^2+\frac{C_2}{\rho^2}+C_3,
\end{equation}
this static scalar potential has been studied in many subjects of physics; for example, in the quantum ring \cite{DantasPLA2015} and also in the Interaction of the magnetic quadrupole moment with an electric field \cite{HassanabadiAP2020}. The potential Eq. \eqref{sspot} is a sample of central potential containing the radial oscillator harmonics and the inverse-square term potentials with a constant term. Moreover, the coefficients $C_1$, $C_2$ and $C_3$ are real constants. Now, by setting the potential Eq. \eqref{sspot} into Eq. \eqref{Schr2}, and also by using a changing variable as $\tilde{s}=\rho^2$, we get
\small
\begin{eqnarray}\label{Schr5}
&&\frac{\mathrm{d}^2\psi_{n\ell}\left(\tilde{s}\right)}{\mathrm{d}\tilde{s}^2}+\frac{1}{\tilde{s}}\frac{\mathrm{d}\psi_{n\ell}\left(\tilde{s}\right)}{\mathrm{d}\tilde{s}}
+\frac{1}{4\tilde{s}^2}\Big[-\left(Q^2C_{m}^2+2mC_{1}\right)\tilde{s}^2\nonumber \\
&&+\left(2m\mathcal{E}-k^2-2QC_{m}\left(\ell-\beta k\right)-2mC_{3}\right)\tilde{s}\nonumber \\
&&-\left(\left(\ell-\beta k\right)^2-2mQ\lambda+2mC_{2}\right)\Big]\psi_{n\ell}\left(\tilde{s}\right)=0.
\end{eqnarray}
\normalsize
As proposed clearly in the first case, we can solve Eq. \eqref{Schr5} through the NU method, such that the eigenfunction can be found as
\begin{equation}\label{WF3}
\psi_{n\ell}\left(\rho\right)=\tilde{N}\rho^{2\xi_{12}}e^{\xi_{13}\rho^2}L^{\xi_{10}-1}_{n}\left(\xi_{11}\rho^2\right),
\end{equation}
where $\tilde{\mathrm{N}}$ is the normalization constant, and also
\begin{eqnarray}\label{alphas2}
\xi_{10}&=&1+\sqrt{2mC_{2}-2mQ\lambda+\left(\ell-\beta k\right)^2}, \nonumber\\ \xi_{11}&=&\sqrt{2mC_{1}+Q^2C_{m}^2}, \nonumber\\
\xi_{12}&=&\frac12\sqrt{2mC_{2}-2mQ\lambda+\left(\ell-\beta k\right)^2}, \nonumber\\
\xi_{13}&=&-\frac12\sqrt{2mC_{1}+Q^2C_{m}^2}.
\end{eqnarray}
Then, the energy eigenvalues associated to the eigenfunction of Eq. \eqref{WF3} can be presented as follows
\begin{equation}\label{energyc2}
\mathcal{E}_{n\ell}=C_{3}+\frac{k^2}{2m}+\frac{QC_{m}}{m}\left(\ell-\beta k\right)+\frac{\tau}{m}\sqrt{2mC_{1}+Q^2C_{m}^2},
\end{equation}
where
\begin{equation}\label{delta2}
\tau=\left[1+2n+\sqrt{\left(\ell-\beta k\right)^2+2m\left(C_{2}-Q\lambda\right)}\right].
\end{equation}
Note that the energy eigenvalue subject to a given collection of parameters.
In this case, considering $C_{2}>Q\lambda$ causes to have a positive quantity in Eq. \eqref{delta2}.
the energy eigenvalues Eq. \eqref{energyc2} is explicitly unbound state if $Q$ and $C_{m}$ have the same sign, $\ell>\beta k$ and also $C_{3}>0$.
Meanwhile, in { Fig.} \ref{Fig1}, we plot energy eigenvalues Eq. \eqref{energyc2} versus the parameter $Q$, according to three different values $n=1,2,3$. We observe that as the value of the parameter $Q$ increases, the value of the energy eigenvalues is almost constant for diagrams $n=1,2$ and is slight increases for diagram $n=3$.

\begin{figure}[htb]
\onefigure[scale=0.55]{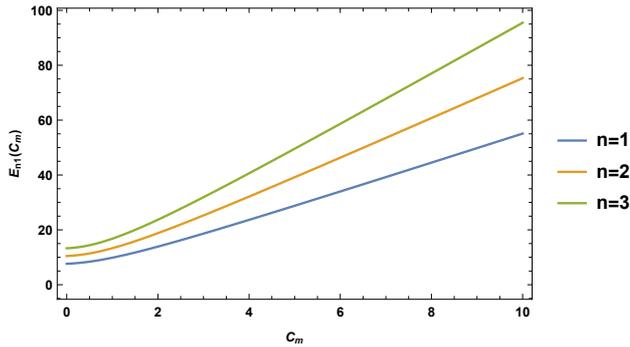}
\caption{The energy eigenvalue $\mathcal{E}_{n\ell}$ from Eq. \eqref{energyc2}, as a function of $C_{m}$ for the values $\ell=\lambda=m=Q=C_{1}=C_{2}=C_{3}=1$, with $k=0.5$ and $\beta=0.5$, according to three different values $\mathrm{n}=1,2,3$.}
\label{Fig2}
\end{figure}

Also, in { Fig.} \ref{Fig2}, we plot energy eigenvalues Eq. \eqref{energyc2} versus the parameter $C_{m}$, according to three different value $n=1,2,3$. We observe that as the value of the parameter $C_{m}$ increases, the value of the energy eigenvalues are increased.

\section{Conclusion\label{Conc}}
In this contribution, we study the quantum dynamics in the background of the interaction of an electric quadrupole moment of a moving particle in an elastic medium with a screw dislocation. Moreover, this medium possesses an electric field and a magnetic field for a non-relativistic particle. the potential energy of a multipole expansion is presented while our focus lies on the electric quadrupole tensor. Also, the Schr\"odinger equation is exactly solved by using the NU method. The interaction in the absence of potential and the presence of a static scalar potential is investigated, such that, we obtain the eigenfunction and energy eigenvalues for these cases.




{
\acknowledgments

The authors thank the referee for a thorough reading of our manuscript and constructive
suggestions.
}



\end{document}